\documentclass{article}
\usepackage{fullpage}
\usepackage{amsfonts,amssymb}
\usepackage{amsthm}
\usepackage{tikz}
\usetikzlibrary{calc}
\usetikzlibrary{arrows}
\usepackage{tikz-3dplot}
\usepackage{color} 
\usepackage[fleqn]{amsmath}
\usepackage{amsthm}
\usepackage{amssymb}
\usepackage{amsfonts}
\usepackage{bbm}
\usepackage{epsfig}
\usepackage{times}
\usepackage{hyperref}
\usepackage{bm}
\usepackage{float} 
\usepackage{appendix} 
\usepackage{lscape}
\usepackage{cite}
\usepackage{mathrsfs}
\usepackage{appendix}
\usepackage{setspace}
\usepackage{mathtools}
\usepackage{mdframed}
\usepackage{authblk}
\theoremstyle{definition}

\begin{document}

\title{Do We Have \emph{Any} Viable Solution to the Measurement Problem?} 

 \author{Emily Adlam  \thanks{The Rotman Institute of Philosophy, 1151 Richmond Street, London N6A5B7 \texttt{eadlam90@gmail.com} }}
\date{\today}

\maketitle

It has become common in foundational discussions to say that we have a variety of possible interpretations of quantum mechanics available to us and therefore we  are faced with a problem of underdetermination. In ref \cite{pittphilsci20537} Wallace argues that this is not so, because several popular approaches to the measurement problem can't be fully extended to relativistic quantum mechanics and quantum field theory (QFT), and thus they can't reproduce many of the empirical phenomena which are correctly predicted by QFT.   Wallace thus contends that as things currently stand, only the unitary-only approaches  can reproduce all the predictions of quantum mechanics, so at present only the unitary-only approaches are acceptable as solutions to the measurement problem. 

Wallace's arguments about the difficulties of extending approaches which are not unitary-only to QFT are quite compelling,  but on the other hand the Everett interpretation and the other extant unitary-only interpretations  have a number of serious epistemic problems which arguably have not yet been satisfactorily resolved (see refs  \cite{AdlamEverett,https://doi.org/10.48550/arxiv.2203.16278}), and thus we are faced with a dilemma: if neither of these obstacles can be overcome then it would seem  that at present   we have \emph{no} viable solution to the measurement problem at all! Or at least, none of the well-studied, mainstream interpretations or modificatory strategies will suffice - and we suspect that many less well-known proposals would also be difficult to extend to non-relativistic quantum mechanics, or would face the same epistemic problems as the unitary-only interpretations. We therefore consider that it remains an urgent outstanding problem to find a viable solution to the measurement problem which can be extended to relativistic quantum mechanics and QFT.

In this article we seek to understand in general terms  what such a thing might look like. We argue that in order to avoid serious epistemic problems, the solution must be a single-world realist approach. We also argue that any single-world realist solution which is able to reproduce the predictions of relativistic quantum mechanics will probably have the property that observable reality does not supervene on dynamical, precisely-defined microscopic beables. Thus we suggest  three possible routes for further exploration:     observable reality could be approximate and emergent, as in relational quantum mechanics (RQM) with the addition of cross-perspective links, or observable reality could supervene on  beables which are not microscopically defined, as in the consistent histories approach, or observable reality could supervene on beables which are not dynamical, as in Kent's solution to the Lorentzian classical reality problem.

We conclude that once all of these issues are taken into account, the options for a viable interpretation or modificatory strategy for quantum mechanics are significantly narrowed down. In light of this fact we have renewed optimism  that it might eventually be possible to arrive at a definitive solution to the measurement problem, although the remaining options still require work before their suitability can be fully assessed.

\section{Background} 

\subsection{The Measurement Problem \label{measurement}} 

We should begin by saying exactly what we mean by `the measurement problem,' and what would constitute a successful solution to it. As we will see later, this is not trivial - in the literature `the measurement problem,' is defined in many different ways, and often these definitions are used to dismiss approaches other than the one favoured by the authors on the grounds that they are not solving the right problem.   We are cautious about this strategy, because from a realist standpoint there must ultimately be some fact of the matter about what the reality underlying quantum mechanics is like, so we don't want to run the risk of ruling out the correct answer by being too exclusionary  - indeed we will later see two  definitions of `the measurement problem' which rule out exactly the set of possibilities which we consider most likely to lead to a correct answer. 
Thus we will attempt here to define the measurement problem in a way that is as inclusive as possible: we will require only that a solution to the measurement problem should render quantum mechanics empirically adequate and susceptible to empirical confirmation (both sine qua non for any viable scientific theory) and should do so without relying on non-physical external observers (for if one is willing to accept non-physical external observers there is no measurement problem at all); we place no further restrictions on what a solution might look like. 
 
Let us elaborate a little. Essentially, the measurement problem arises from the fact that  unitary-only quantum mechanics does not seem to describe a unique macroscopic world: in general it gives rise to a large number of superposed macroscopic possibilities, looking nothing like the unique observable  reality that we experience.  Thus, on the face of it, unitary-only quantum mechanics is not empirically adequate, since it fails to predict any specific outcomes for measurements at all. The measurement problem, then, can be regarded as the problem of showing how to extract actual predictions out of this formalism. Famously, this can be done by simply assuming that  whenever an observer performs a measurement the wavefunction undergoes a non-unitary collapse to yield a single measurement outcome with probabilities in proportion to the mod-squared amplitudes. But this approach requires us to treat observers as entities external to the theory which can't be understood as a part of ordinary physical reality, and although this may be acceptable to operationalists or idealists, it is untenable for those of a more realist, physicalist persuasion. 

Thus we argue that in order to make quantum mechanics empirically adequate and susceptible to empirical confirmation on terms acceptable to realists, the formalism must be supplemented or reinterpreted to achieve at least the following three things: 

\begin{enumerate}
	
	\item The theory must provide a mechanism by which the observable   world(s) perceived by observers can arise, without treating agents or measurements as external to the theory
	
		\item The mechanism by which the observable world(s) arise must ensure or make probable that the world(s) exhibit relative frequencies close to the quantum mod-squared amplitudes
		
		\item The  observable world(s) perceived by  observers must be systematically related to each other in some way
	 
\end{enumerate}

Criteria 1) and 2) are clearly necessary if we want the theory to be empirically adequate,  and have been widely discussed in the literature. Criterion 3) has not received so much attention, but we would argue that it is also very important. For in order to empirically confirm a theory we must make certain assumptions about the reliability of our evidence for that theory - at minimum we must assume that the records  perceived by an observer at a given time  are usually related in some systematic way to the actual macroscopic events witnessed by previous versions of that observer, and since in practice empirical confirmation also relies on reports by other observers, we must additionally assume that the observable world perceived by one observer is usually related in some systematic way to the observable worlds perceived by other observers. Thus any viable interpretation of the resulting theory should entail that these assumptions are largely accurate, in order that the theory can be susceptible to empirical confirmation by the usual methods of science (this argument is made in more detail in ref \cite{https://doi.org/10.48550/arxiv.2203.16278}). 

 Now, what we have said about the measurement problem and the criteria for its successful solution may look a little different to some popular accounts. In particular, it is common to suggest that  the measurement problem is related to the need to provide an ontology for quantum mechanics\cite{Esfeld_2020} -  but in our view this is somewhat misleading. Certainly, one way of solving the measurement problem would be to provide an underlying ontology and show how `measurements' and other macroscopic events arise naturally from this ontology in such a way as to satisfy the criteria above, but this is not the only possible approach. For example, another way to solve the measurement problem would be to add to the quantum formalism a piece of mathematics which ensures in a natural way that during the process of decoherence, branches of the wavefunction not only become orthogonal, but also all but one of them ceases to exist, with the probability for a branch to survive roughly proportional to its quantum mod-squared amplitude. This addition to the formalism would ensure that decoherence produces exactly one reality shared by all observers,  and thus in our view  it would  count as a solution to the measurement problem,  even if we were not able to come up with an ontological interpretation for it. So although it would undeniably be satisfying to have a concrete ontology for quantum mechanics, ontology is not necessarily crucial to the measurement problem - and therefore the measurement problem is still relevant for structural realists and other kinds of realists who prefer not to  make strong ontological claims, since the problem persists even if one is not concerned about ontology.\footnote{  We note that there are a class of ideas sometimes mentioned in connection with the interpretation of quantum mechanics, like retrocausality\cite{Price2012-PRIDTI-2} and superdeterminism\cite{10.3389/fphy.2020.00139}, which  are not in and of themselves solutions to the measurement problem: they don't say anything in particular about the emergence of our shared observable reality, rather they are simply properties that a solution to the measurement problem may or may not have (for example, the transactional interpretation\cite{Cramer,RK} is a solution to the measurement problem which has the property of being retrocausal). So we will not have much to say about these ideas in this article - not because there is anything wrong with them, but because they are not answers to the questions we are addressing here.}

\subsection{  Unitary-Only and Single-World Realist Approaches \label{unitary}} 

We begin by delineating two important classes of (putative) solutions to the measurement problem. First we have  `unitary-only' approaches - that is, approaches which do not add anything to the standard formalism of unitary quantum mechanics, so there are no wavefunction collapses, no additional classical variables, additional classical trajectories or anything of that nature. This class includes the Everett interpretation\cite{Wallace}, as well all of the `orthodox interpretations, (see refs \cite{https://doi.org/10.48550/arxiv.2203.16278,cuffaro2021open}) -  some versions of the Copenhagen interpretation\cite{Bohr1987-BOHTPW,Heisenberg1958-HEIPAP,pauli1994writings}, neo-Copenhagen interpretations\cite{Zeilinger1999-ZEIAFP,Zeilinger2002,brukner2015quantum,articlebanana,demopoulos2012generalized,Janas2021-JANUQR,2004neoc}, QBism\cite{QBismintro}, some pragmatic interpretations\cite{Healey2012-HEAQTA} and some versions of relational quantum mechanics\cite{1996cr}.

Second we have the set of `single-world realist' (SWR) approaches -  approaches which maintain that there exists a unique observable reality which different observers by and large agree on. That is to say, although of course no observer sees everything and observers may sometimes make mistakes, nonetheless there exist reliable mechanisms by which any pair of observers can in principle come to agree on the content of observable reality (and in particular, on the outcomes of measurements). So in these approaches we don't have  observers living in different Everettian branches who can no longer interact, and we don't have observers confined to distinct perspectives which never come into contact. Prominent SWR approaches include the de Broglie-Bohm approach\cite{holland1995quantum} and spontaneous collapse approaches\cite{Bassi,Tumulka_2006,Gisin2013}. Note that these approaches are often described as `interpretations,' but we will continue using the term `approach', because SWR approaches typically alter the standard formalism of unitary quantum mechanics in some way and therefore they are really modificatory strategies rather than interpretations.

Now, these two categories aren't exhaustive - there can be approaches which are neither unitary-only nor SWR. For example, consider a `many-minds' approach\cite{Lockwood1996-LOCMMI-2}, which is simply the Everett interpretation with the addition of a set of `minds' such that different minds go into different branches during branching events. Evidently this approach is not unitary-only since the set of minds is added to unitary quantum mechanics, and nor is it SWR since different minds end up trapped within isolated branches. However, in our view the motivation for approaches like this is questionable - if one is going  to add something to unitary quantum mechanics, then one might as well add something which allows us to avoid the epistemic problems associated with non-SWR approaches -  so we will henceforth largely disregard this domain of possibilities.
 
 Conversely, the two categories don't appear to be mutually exclusive, so one might imagine that there   could be an approach which is both unitary-only and also SWR. However, the fact that all existing unitary-only interpretations fail to be SWR   is not incidental; it is a natural consequence of structural features of unitary quantum mechanics. Specifically, unitary quantum mechanics does not provide a mechanism for singling out a unique outcome to a measurement for any one observer. So in a unitary only-approach either we must say that each observer sees more than one outcome of a measurement, so we end up with distinct branches of reality containing observers who can't interact; or  if we insist that each observer sees only one outcome, then since unitary quantum mechanics doesn't say anything about how  an  observer comes to see some particular outcome, an approach which adds nothing to unitary quantum mechanics can't possibly postulate any systematic  relations between the measurement outcomes seen by different observers, so the   observers can't exchange information and thus  they can't reliably come to agreement on the content of observable reality.   Therefore  it seems likely that something must be added to unitary quantum mechanics if we want to arrive at a SWR approach.

Refs \cite{AdlamEverett,https://doi.org/10.48550/arxiv.2203.16278} argue that non-SWR approaches face serious epistemic problems which undermine their claim to be successful interpretations of quantum mechanics. Specifically, ref \cite{AdlamEverett} argues that in the context of the Everett interpretation, the empirical confirmation for quantum mechanics is undermined because in order to make sense of empirical confirmation the Everettians need to assume that   we are inside a branch of the wavefunction which is associated with a high mod-squared amplitude, but the theory does not have the resources to justify that assumption. If this is correct, the Everett interpretation cannot fulfil the second requirement for a successful solution to the measurement problem as stated in  section \ref{measurement}, Similarly,  ref \cite{https://doi.org/10.48550/arxiv.2203.16278} argues that in the context of an orthodox interpretation, the empirical confirmation for quantum mechanics is undermined by the fact that the orthodox interpretations have the consequence that observers can never learn anything about the experiences of other observers or even past versions of themselves. If this is correct, the orthodox interpretations cannot fulfil  the third requirement for a successful solution to the measurement problem as stated in section \ref{measurement}.   Moreover, these problems are also not incidental - they are natural consequences of the fact that by definition, in non-SWR approaches there is no unique observable reality which observers by and large agree on. The absence of a shared observable reality inevitably leads to problems, since the assumption that observers  are all observing roughly the same reality plays a crucial role in the epistemology of science. 

Therefore we believe that a solution to the measurement problem which can fulfill all three of the criteria in section \ref{measurement} is  likely  to be a SWR approach - or at least \emph{approximately} a SWR approach, an idea which we will shortly explore in more detail. We reinforce that this motivation for pursuing SWR approaches is  nothing to do with wishing to preserve na\"{i}ve classical intuitions about the nature of reality: the goal is simply to have an account of  quantum mechanics which allows that the theory is empirically adequate and susceptible to empirical confirmation, something that must be achieved by any  viable interpretation or modificatory strategy.

\section{Quantum Field Theory \label{relativistic}} 

The predictions of relativistic quantum mechanics and QFT, understood in purely operational terms,  have been confirmed to an extremely high degree of accuracy, and thus no interpretation or modificatory strategy for quantum mechanics which is incapable of being extended to QFT is empirically adequate. But SWR interpretations have historically struggled to accommodate QFT, and moreover, as Wallace argues\cite{pittphilsci20537}, this is not just bad luck: there are structural issues that make   relativistic quantum mechanics quite hard to reconcile with existing SWR approaches. 

The origin of the problem is that in order to have an SWR account, we must insist that one of the possible histories encoded in the unitary evolution of the quantum state is physically real, and the others are not real (they represent mere possibilities, or possibly some form of modal or nomic structure). And the distinction between `physically real' and `mere possibility' surely cannot be approximate or emergent: reality is not the kind of thing which comes in degrees. Yet the macroscopic histories encoded in the unitary evolution of the quantum state are only approximate and emergent, since they become distinct from one another via the mechanism of decoherence, which itself is approximate and emergent. So SWR accounts typically have to proceed by adding to unitary quantum mechanics something that is not approximate or emergent, in order that our  observable reality can supervene on it  - as Wallace puts it, this requires us to find a `\emph{microphysically-stateable, precisely-defined dynamical variable which, on coarse-graining and restriction to the non-relativistic particle-mechanics regime, nonetheless delivers coarse-grained particle position or some appropriate surrogate}'\cite{pittphilsci20537}.  This strategy is made explicit in the `primitive ontology' approach, advocated in particular by Allori, Goldstein, Tumulka and Zanghi\cite{pittphilsci9337,10.2307/40072290,pittphilsci11651}, which enjoins us to make sense of quantum mechanics by identifying some ontology which `\emph{lives in three-dimensional space}' and `\emph{constitutes the building blocks of everything else}'\cite{pittphilsci11651}.

It is quite straightforward to identify a suitable primitive ontology as long as we confine ourselves to non-relativistic quantum mechanics - for example, the corpuscle positions play this role in the de Broglie-Bohm approach. But as soon as we move to quantum field theory, primitive ontology approaches are in trouble. For no one has yet found a way to define within quantum field theory a precisely-defined, microphysically-stateable dynamical variable which gives rise to all of the appropriate macroscopic histories in the macroscopic limit; and Wallace further argues that   general features of QFTs make it implausible that there exists any variable which has all of these properties. For in QFT the underlying microscopic theory shares almost no features at all with the classical world that emerges from it - there are no particles, no classical field states, and even the parameters have different values, since the classical parameters are obtained from the bare microsopic parameters by renormalisation. As Wallace puts it, `\emph{the relation between the ‘fundamental’ and the empirically-relevant in quantum field theory is complicated,
	indirect, dynamically mediated and cutoff-dependent. It is very hard to see how
	this could be made compatible with the primitive-ontology approach, or indeed
	with any approach committed to a description of a theory’s empirical content
	directly in its microphysical vocabulary}' \footnote{Incidentally, this point illustrates why older versions of the Everett interpretation, which postulate precisely defined worlds or minds in addition to the wavefunction, must inevitably yield to the modern emergence picture\cite{Everett}: as soon as the Everettians provide a prescription to make their worlds precise, they lose their most crucial advantage.}.

Wallace illustrates this point by examining existing attempts to generalise SWR approaches to QFT. For example, the de Broglie-Bohm interpretation has some relativistic generalizations, where we replace the primitive ontology of `corpuscle positions' with another quasi-classical variable: for example, ref \cite{BohmQFT} uses a variable encoding both the number of particles and the positions of those particles, ref \cite{Struyve_2007} uses the electromagnetic field, and ref \cite{Valentini:1992bxa} uses Grassman fields to model fermionic fields. But none of these approaches works in the most general case because particle number, particle position and field states are all emergent in QFT: they  appear only in certain regimes at certain energy scales, and they look different or disappear completely as we move through energy scales. Similarly, spontaneous collapse models have some relativistic generalisations\cite{Tumulka2006,tumulka2020relativistic,PhysRevResearch.1.033040}, but they are defined only for systems described in terms of particles, so they will likewise not work in all regimes of QFT. Thus it would seem that none of these approaches can be precisely defined at  the most fundamental level, and therefore it's unclear how  their primitive ontology could suffice to pick out a unique physically real observable world.

 Of course it's possible that the apparent nonexistence of a suitable   primitive ontology within QFT is just down to a collective failure of imagination and  someone will eventually discover a sufficiently classical set of dynamical microscopic variables which can be precisely defined in all regimes of QFT. But we share Wallace's pessimism on this point, and thus we will henceforth assume that this cannot be done.  Likewise, it is possible that there is some way to solve the epistemic problems for unitary-only approaches that we described in section \ref{unitary}, but for the sake of argument we will henceforth suppose that there is not. This leaves us with an interesting problem: if neither the unitary-only approaches nor the well-known SWR approaches like de Broglie-Bohm and spontaneous collapse are viable solutions to the measurement problem, what possibilities remain?

Consider again the following four desiderata for the beables on which the observable world is to supervene, which according to Wallace cannot all be realised within relativistic quantum mechanics: 

\begin{enumerate} 
	\item Microphysically-stateable
	
	\item Precisely-defined
	
	\item Dynamical
	
	\item Empirically adequate (i.e. delivers appropriate coarse-grained macroscopic histories  in some appropriate limit)
	\end{enumerate}

Can we afford to lose any of these properties? `Empirically adequate' seems non-negotiable: we have not solved the measurement problem if we cannot reproduce the predictions of quantum mechanics at least in some appropriate limit. But the others are more questionable. The requirement that the beables should be precisely-defined may at first seem very natural, since `the real world' is to supervene on these beables and `the real world' surely cannot be approximately defined - but recall that our reasons for demanding a SWR approach are not concerned with preserving classical intuitions about `the real world' but rather with solving epistemic problems, and there is no obvious reason why a solution to these epistemic problems cannot be somewhat approximately defined. So we could perhaps compromise on `precisely-defined.' And it also doesn't seem obvious that either `microphysically-stateable' or `dynamical' is indispensable: the idea that the fundamental beables must be `microphysically-stateable' is a product of the  reductionist paradigm, while the idea that they must be `dynamical' reflects the time-evolution paradigm, and we have argued previously\cite{Adlamspooky, adlamfund} that both of these paradigms should be re-examined and possibly given up.

These observations suggests that a successful SWR approach to quantum mechanics will probably not postulate a `primitive ontology' in the intuitive sense of that term - the shared observable reality it postulates may supervene on variables which are non-dynamical or non-microscopically defined, or which are not \emph{precisely} defined at any scales. Of course, evidently any empirically adequate theory must postulate \emph{something} located in spacetime, even if only macroscopic reality itself, so of course there will always be some kind of ontology which lives in three-dimensional space, but if this ontology is non-dynamical, not microscopically-defined or not precisely-defined, it's unclear that it can sensibly be regarded as playing the role of   `building blocks' in the way suggested by the primitive ontology approach. Indeed, this observation reinforces the importance of being somewhat circumspect in stating what exactly counts as a solution to the measurement problem: if one insists, as for example in ref \cite{Allori2013-ALLOTM-2}, that a solution to the measurement problem requires a theory in which  `\emph{physical objects are constituted by a primitive ontology}'  (indeed, ref \cite{pittphilsci11651} makes the even stronger claim that the primitive ontology approach `\emph{tells us what structure proper fundamental physical theories ought to have}') then one runs the risk of allowing only solutions which cannot possibly reproduce QFT.

\subsection{Further Criteria \label{criteria}}

We will shortly discus the three possible routes to a solution that we identified above in more detail, but first we propose some further features which we consider likely to appear  in any solution to the measurement problem which is capable of reproducing the predictions of QFT.
 
\paragraph{Preserve unitary quantum mechanics}

Mainstream SWR approaches often make quite significant additions to unitary quantum mechanics, such as adding dynamical collapses or introducing quasi-classical particles. But there are good reasons to think that a successful interpretation of quantum mechanics should probably leave the framework of unitary quantum mechanics very substantially intact, at least at the microscopic level. For unitary quantum mechanics has achieved extraordinary empirical success as a description of the microscopic world, and it is very difficult to change the framework in any significant way without sacrificing a significant part of that empirical success - this is precisely the reason why QFT generalizations of de Broglie-Bohm and GRW have been so hard to come by. So although we may of course choose to reinterpret the framework of unitary quantum mechanics, re-imagining the quantum state as an element of nomic or modal structure or  a mere predictive tool, it seems likely that a correct solution to the measurement problem will preserve it largely intact at least as a functional structure: the unitary evolution should not be interrupted by collapses, and we should not add quasi-classical variables in regimes which according to the theory are  highly non-classical. This suggests that  any beables we add to quantum mechanics should probably be largely epiphenomenal, at least with respect to unitary quantum mechanics: presumably they will be at least partly determined by the evolution of the underlying quantum state, but they should have no substantial effect on the evolution of the underlying quantum state.

\paragraph{ Make full use of decoherence} 

An additional problem with existing SWR approaches is that they sometimes fail to give decoherence its due.  Decoherence refers to the process by which quantum systems become coupled to their environment, which leads to the distinct branches of quantum superpositions becoming orthogonal and non-interacting\cite{sep-qm-decoherence}. Evidently decoherence looks a great deal like the  emergence of classicality - it provides us with quasi-classical, non-interacting branches over which we can define a classical probability distribution, so the only thing missing is some mechanism to select and actualise just one of those branches. Indeed, decoherence is so close to what we need to explain the appearance of a macroscopic classical reality that in our view it is simply not credible that the existence of decoherence could be a complete coincidence which has nothing to do with the emergence of classicality. Thus putative solutions to the measurement problem which put the classicality in directly at the fundamental level by introducing a set of microscopically-defined quasi-classical variables seem poorly motivated, because they are ignoring the fact that there is already a well-understood mechanism to produce quasi-classical variables in the macroscopic limit. 

This is perhaps clearest in the case of approaches incorporating a literal collapse of the wavefunction, such as the GRW picture\cite{GRWlett,Tumulka_2006}. In this model, classical reality appears as a result of the collapse of the wavefunction, which selects out one branch of the wavefunction and suppresses all the others. Moreover, the collapse is spontaneous, i.e. it is nothing to do with decoherence, so decoherence mostly doesn't get a chance to occur, and even if it does occur it is irrelevant, - we don't need decoherence to make the branches non-interacting, because only one branch will survive in any case. So in our view these collapse-based approaches are doing too much: they are not taking advantage of the fact that unitary quantum mechanics already has a well-defined process by which classicality can emerge.

The de Broglie-Bohm approach\cite{holland1995quantum} fares better: decoherence does play a role in the emergence of classical reality in this picture, because it is decoherence which ensures that after a macroscopic event like a measurement, the branches become separate and non-interacting and therefore the corpuscles will not subsequently be influenced by branches other than the one they ultimately ended up in. However, decoherence is not fully taken advantage of here - the only important feature of decoherence in this picture is that it ensures branches don't interfere, and the fact that it also yields what look a lot like a set of quasi-classical worlds is irrelevant, since the de Broglie-Bohm approach provides us with a completely independent set of fundamental quasi-classical variables  on which the classical world ultimately supervenes. So the de Broglie-Bohm approach also appears to be doing too much. 

Moreover, the fact that these approaches don't make full use of decoherence is part of the difficulty in generalising them to QFT -  because they do not allow  our quasi-classical reality  to emerge naturally in the process of decoherence, some suitable microscopically-defined quasi-classical variables must be found within QFT on which classical reality can supervene, and it turns out to be quite hard to find such variables. On the basis of these examples, it seems likely that any quasi-classical beables we add to quantum mechanics should be in some sense defined with reference to decoherence, so they can take advantage of structures that already exist in unitary quantum mechanics.

\paragraph{Be flexible}

As   noted earlier, many approaches to solving the measurement problem explicitly have the goal of providing an ontology for quantum mechanics. But that is exactly what gets them into trouble, because, as Wallace argues in ref \cite{pittphilsci15292}, quantum mechanics can be applied to a very large variety of different types of systems and it is not at all clear that there is any one ontology that makes sense for every application. In particular, a number of popular approaches are based on an ontology of particles (this is true for de Broglie-Bohm, and also for most relativistic formulations of GRW), and yet  quantum mechanics describes many things other than particles. We have already noted that this causes serious problems if one tries to apply these approaches to regimes of QFT in which particles are only emergent and approximate, but more generally it can cause problems even within standard non-relativistic quantum mechanics when applied to systems which have no natural `particle' description, such as collective excitations of solid bodies or internal degrees of freedom\cite{pittphilsci11651}.  This suggests that anything we add to quantum mechanics should not depend too sensitively on the ontology of the system being described -  in Wallace's terms, we need an `\emph{interpretive recipe}' which `\emph{tells us, for any given quantum theory, how to understand \textbf{that} theory.}'

\section{The Emergence Approach}

We now move to a more detailed examination of our three suggested routes to a QFT-friendly, SWR solution to the measurement problem. The first possibility  involves giving up the assumption that the unique observable reality associated with a SWR approach must be precisely defined. Of course, if one wants a `unique real world' purely on the grounds of metaphysical convictions about the nature of reality, then it seems clear that what is `real' cannot be approximate and emergent. But our reasons for requiring a unique real world are not metaphysical but \emph{epistemic}, and therefore it is conceivable that the `unique real world' could in fact be somewhat approximate - it need only be definite enough to address the epistemic difficulties faced by non-SWR approaches. In particular,   we require that the  observable realities perceived by macroscopic observers satisfy the three conditions set out in section \ref{measurement} - but  since on a physicalist conception macroscopic observers themselves are somewhat approximate and emergent, there is no reason to think that the observable reality they perceive should be defined completely precisely, and thus there is no reason to think  the three requirements set out in section  \ref{measurement} must be satisfied completely precisely. 

The  idea of a \emph{vaguely defined}   observable reality  could be cashed out in several different ways - in particular, the vagenness in question could be semantic vagueness,  where there is some ambiguity about the referent of a term, epistemic vagueness, where there is some vagueness in our knowledge of the world, and ontic vagueness, which refers to vagueness in the world itself, i.e. vagueness that would persist even if we had perfect knowledge and completely precise terms\cite{Barnes2010-BAROVA-2}. Epistemic vagueness doesn't seem particularly helpful here, because if reality is precisely defined and it is merely our knowledge of reality that is vague, then we still have the problem of extracting  a precisely-defined  reality out of the approximately-defined quantum histories. Ontic vagueness seems more promising - perhaps we could say that there exists one shared observable reality but it has vague boundaries and thus there are no completely precise facts about what occurs in it.  To our knowledge this is not an approach which has been explored within physics, and indeed it would pose unique challenges, because there are good reasons to think that ontic vagueness cannot be faithfully written in the language of mathematics\cite{Chen_2022}, so we might need to develop new tools in order to write down a theory of physics with genuine ontic vagueness. There has been some interesting recent philosophical work in this space - for example, Chen has argued for the possibility of fundamental laws of nature which exhibit genuine vagueness\cite{Chen_2022}, so perhaps we could imagine invoking fundamentally vague laws in order to generate an ontically vague macroscopic world. 

However, in  this section we will largely focus on  \emph{semantic} vagueness - that is, we will suppose that the facts about what there is in the world are perfectly well defined, but there is some vagueness about what  the term   `observable reality' refers to. Thus since we don't need to be able to state precisely what the shared observable world contains, we can potentially identify it with one of the approximately-defined classical histories that emerge from decoherence.  But decoherence on its own doesn't single out one particular macroscopic history or offer any mechanism which would allow macroscopic observers to reach agreement on the content of classical reality, so we still need to add such a mechanism. In principle this could also be defined in an approximate way - the empirical confirmation attached to quantum mechanics depends only on the assumption that observers can generally come to agreement on  macroscopic events like measurement outcomes, so it's all right if they sometimes differ on minor facts about the fuzzy edges of classical reality - but even so it can't come out of unitary quantum mechanics, because    we saw in section \ref{unitary} that   unitary quantum mechanics doesn't  provide a route to a shared observable reality, even an approximate one.  So what is the minimal addition we can make to unitary quantum mechanics to pick out, at least approximately, a  classical reality that can be shared by multiple observers?

We will use relational quantum mechanics (RQM) as a representative example of a unitary-only approach to quantum mechanics, although a similar approach could potentially be taken within other orthodox interpretations. The original formulation of RQM\cite{1996cr} was based on the idea that events always happen relative to physical systems: during an interaction between two physical systems $A$ and $B$, we get a `quantum event' in which some variable of system $S$  takes on a definite value $V$ relative to system $A$. If $A$ is a macroscopic observer, Alice, then this event can be interpreted as Alice measuring the system $S$ in some basis and obtaining an outcome $V$. But RQM tells us that the value $V$ is only correct   relative to $A$ - relative to other systems it is still correct to describe $A$ and $S$ by a  unitary evolution  in which the interaction simply results in them becoming entangled  so the variable $V$ doesn't take on a  definite values.  This version of RQM ensures that each observer sees a single definite outcome for every measurement, but it is not a SWR approach according to our definition because it tells us that all facts are relative to observers and hence it is meaningless to even ask whether two observers agree about some measurement outcome, so there can't be a mechanism for macroscopic observers to reach agreement on the content of classical reality. 

To address this problem, ref \cite{https://doi.org/10.48550/arxiv.2203.13342} proposes adding to RQM a postulate called `cross-perspective links' which stipulates that whenever a quantum event occurs in which systems $A$ and $S$ interact and variable $O$ takes on the definite value $V_A$ relative to $A$, then provided that $A$ does not undergo any interactions which destroy the information about $V$ stored in $A$'s physical variables, if $A$ and $B$ subsequently interact in such a way that the physical variable $O'$ representing the information obtained by system $A$ about the variable $O$   takes on the definite value $V_B$ relative to  $B$, then $V_A$ will match $V_B$. In the case where $A$ and $B$ are macroscopic observers Alice and Bob, this means that if Alice measures $S$ and then Bob performs a measurement on Alice which is expected to provide information about her measurement result (for example, he could simply ask her which result she got) then the result of Bob's measurement will match the result of Alice's.  This postulate entails that as an epistemic community of observers interact they will build up a shared observable reality composed of a large number of variables whose values all the observers agree on; in ref \cite{pittphilsci19401} Healey offers a detailed account of the way in which such epistemic communities can arise within unified `decoherence environments'\footnote{The postulate of cross-perspective links  may sound somewhat ad hoc, but it can to some degree be justified on the grounds that RQM is intended to be a physicalist appraoch and thus Alice's conscious awareness of seeing a definite value of some variable must  supervene on some physical variable of Alice which should be accessible to Bob by the right kind of measurement. Ultimately we might hope to have a more constructive account of the nature of this variable and the way in which Bob accesses it, but the postulate at least serves to indicate what kind of structure must be added to an orthodox interpretation if we are to arrive at a SWR approach.}.

RQM with cross-perspective links exhibits exactly the kind of semantic vagueness we were looking for: interactions between members of an epistemic community result in the community eventually sharing a single `shared observable reality' but since the cross-perspective links postulate still still allows observers to disagree about the values of variables when that information has not yet been disseminated through the epistemic community, the referent of that term `shared observable reality' is not completely well-defined.   Indeed, although decoherence combined with cross-perspective links together guarantee that any observers who come into causal contact will belong to the same epistemic community, in principle there could be other observers out there belonging to epistemic communities disconnected from ours  who share a  completely different  observable reality, and thus strictly speaking this is not really a `single-world realist' approach. But nonetheless it is single-world realist enough to solve the epistemic issues which motivated us to seek a SWR view, for after all, if there are aliens out there somewhere in a galaxy far far away, the empirical confirmation  attached to quantum mechanics surely does not depend on any claims about what those aliens have observed: what matters is that the community of human observers who have together arrived at the theory of quantum mechanics are correct in their belief that they are by and large all sharing the same classical reality.

RQM with cross-perspective links also satisfies the criteria set out in section \ref{criteria}:

\begin{enumerate} 

\item It tells  us that unitary quantum mechanics is always correct - for example, even if Alice measures a particle and records an outcome $O$, nonetheless with respect to Bob, Alice and the particle remain in the entangled state predicted by unitary quantum mechanics applied to their joint system, so if Bob has good enough experimental technique he can in principle perform interference experiments or detect this entanglement by means of measurements in different bases. The cross-perspective links postulate does not change this, because it adds nothing except a stipulation about the relationship between the outcomes obtained by two different observers - a topic upon which unitary quantum mechanics itself is conspicuously silent.

\item It tells us that decoherence is responsible for the emergence of classicality, since it is decoherence which provides   a preferred basis and thus ensures that macroscopic observers perceive measurements as having a single outcome in one basis, rather than having a single outcome in each one of a range of possible bases (see ref \cite{https://doi.org/10.48550/arxiv.2203.13342} for details). 

\item It is quite flexible,  since the values of a given system can take values relative to any other physical system whatsoever, and therefore the prescription can be applied to any quantum description which involves distinct physical systems interacting, which includes a very wide range of quantum descriptions (although as we will shortly see, there may be some regimes of QFT which don't fulfil this requirement).

\end{enumerate}

Naturally there are still issues to be worked out with this approach. In particular, as discussed in ref \cite{https://doi.org/10.48550/arxiv.2203.13342}, in the case of a quantum event arising from the interaction between  two \emph{microscopic} quantum systems $A$ and $B$, the interaction Hamiltonian will not single out a unique basis of measurement, and thus the corresponding quantum event involves the realisation of not just one but a range of different measurement outcomes, one for every possible basis of measurement. So we will have to think of $A$ and $B$ as `measuring' one another in a range of different bases, and cross-perspective links will have to be applied to every one of $B$'s bases which coincides with one of $A$'s bases, such that in the limit as $A$ and $B$ become the macroscopic observers Alice and Bob we can be assured that there will be a unique outcome shared by both Alice and Bob in the preferred basis singled out by decoherence. One  might worry that this leads to a theory with a lot of superfluous structure: we end up defining a complex network of shared outcomes between an enormous number of microscopic particles, even though most of those outcomes will be undetectable and irrelevant for future evolution. Arguably it would be preferable for `cross-perspective links' to be realised only in the macroscopic limit as interactions become recognisably measurements in a single basis, but it's not straightforward to see how to achieve that, because the definition of cross-perspective links as used in ref  \cite{https://doi.org/10.48550/arxiv.2203.13342}  seems to require precisely defined systems and interactions. 

This points to a further problem - although RQM does a good job of preserving the structure of unitary quantum mechanics, it's still not obvious that the approach can be generalized to QFT. This is because RQM leans heavily on the existence of well-defined systems and interactions - which is very reasonable within  non-relativistic quantum mechanics but is difficult to maintain within relativistic quantum mechanics, where particles and trajectories are merely emergent. In Sorkin's words, `\emph{ the concept of “subsystem” takes center stage in many interpretations of the quantum formalism. In ``chemistry” (the theory of nuclei, electrons, and Coulombic interaction), one could perhaps construe a subsystem as a definite collection of particles, but what could it mean in the context of quantum field theory in curved spacetime, say?'}\cite{Sorkin_2007} In the original formulation of RQM we could perhaps accept that `systems' and `interactions' are only approximately defined and emergent, but this is more difficult in the SWR version with cross-perspective links, since the definition of cross-perspective links requires precisely defined systems and interactions. 

One possible option here would be to understand the `systems' in RQM in a different way - i.e. rather than thinking of a `system' as a particle or collection of particles, we could understand a `system' as a region of space, or spacetime\footnote{Thanks to Carlo Rovelli for this suggestion.}. This would lead to a picture where spacetime itself is the bearer of the relational descriptions - events occur relative to some region of spacetime, so each way of separating spacetime into parts would correspond to a different relational description. One advantage of this approach is that it straightforwardly answers questions about the identity of a `system' over time - no region of spacetime persists over time, so  no `system' does either, but we can still have an analogue of identity over time in terms of   cross-perspective links relating the descriptions relative to spacetime regions standing in certain relations to one another.\footnote{Quantum gravity is not the main topic of this article, but it's worth nothing that   if we take this approach one might think the same problem would re-emerge when we move to quantum gravity - many physicists believe that spacetime itself is only approximate and emergent in quantum gravity, so if `systems' are defined in terms of regions of space then systems are still approximate and emergent when we get to quantum gravity.   However, perhaps there might be a way to understand a `system' as some subregion of whatever substratum underlies spacetime in quantum gravity (e.g. a single node or set of nodes in a spinfoam) with these subregions becoming regions of spacetime in an appropriate limit}.

Another possible option would be to formulate cross-perspective links in a way that works even if the systems and interactions involved are only approximately defined.  In principle this seems reasonable, as we have already noted that `epistemic communities' don't need to be sharply defined and hence the links between them also shouldn't need to be sharply defined. Indeed, the sharp way in which cross-perspective links is defined is itself a potential problem, because one might worry that in realistic situations when $A$ and $B$ are macroscopic observers Alice and Bob, it's unlikely that Bob will measure Alice in the \emph{exact} basis which corresponds to the outcome of her measurement on $S$: after all, if we assume that spacetime is continuous and hence there exists a continuum of possible bases, then the probability that he measures in exactly the same basis is zero. So it seems likely that cross-perspective links must in any case be made a little more approximate -  perhaps we should think of Alice and Bob as measuring in a small range of possible bases, since Alice and Bob themselves are  only approximate and emergent, and then we could generalize the cross-perspective postulate such that it applies any time there is non-zero intersection between the range of bases employed by the two observers\footnote{Or we could insist that spacetime is in fact discrete and hence there is not a continuum of possible bases, in which case it is more plausible that Bob could manage to reliably measure Alice in the exact basis corresponding to the outcome of her measurement. In fact RQM is often linked with loop quantum gravity, which does indeed tell us that spacetime is discrete, so this route could make sense.}.

Alternatively, we could attempt to add to RQM some ontology which \emph{is} sharply defined. For example, ref  \cite{https://doi.org/10.48550/arxiv.2203.13342}  postulates a set of objective quantum events distributed across spacetime; the unitary quantum description, including all the mechanics of QFT, can then be understood as a guide for navigating this set of events by telling us the probabilities for future events conditional on some set of known past events. Because the events are determined only stochastically by the underlying quantum description, they can in principle be sharply defined even though the quantum description does not always provide us with sharply-defined systems and interactions, so the postulate of cross-perspective links can be understood as applying directly to the precisely-defined quantum events rather than to the approximately-defined systems and interactions featuring in the unitary quantum description. 

 However, the version of RQM incorporating a stochastic distribution of events looks a little like the relativistic version of the GRW collapse models\cite{Tumulka2006,tumulka2020relativistic}, which likewise postulates a stochastic distribution of events across spacetime, so one may worry that this version of RQM will face the same difficulties with regards to being generalized to QFT. Thus more must be done before we can be confident that this version of RQM will work for QFT - in particular, we might want to know more about how to define the overall distribution of the events, given that RQM tells us the  unitary quantum description can only be understood from the perspective of some particular location within the network of quantum events, and thus unitary quantum mechanics does not describe the distribution of events as a whole. In fact, it seems likely that if we take this route we are left with the problem of  showing how to extract from the quantum description some `higher-level' non-microscopically-defined beables, and therefore this route ends up looking similar to the non-reductionist approach, so let us now turn to that approach.

\section{The Non-Reductionist Approach \label{conshit}  }

The second route we will explore involves allowing that the unique real world associated with a SWR approach could supervene on beables which are not microscopically defined. This idea may initially seem very counterintuitive, but in fact it seems like quite a good fit for QFT given that QFT is an effective field theory, which is to say it describes physics at a certain scale while ignoring details of the physics at smaller scales.  The reason this is possible is because of a phenomenon known as universality\cite{https://doi.org/10.48550/arxiv.1406.4532}: the way that theories change as we move through different scales is described by the renormalisation group, and a crucial property of the   renormalisation group is that it takes many different underlying theories  to the same fixed point at higher scales, which is to say that many of the  details of physics at very small scales make no difference to  physics at the scales we are able to observe.  In this sense, as Wallace notes, `\emph{A QFT is probably best understood as an equivalence class of theories reproducing the same large-scale particle physics phenomenology}'\cite{pittphilsci15292}. This scenario seems quite friendly to a non-reductionist approach to quantum mechanics, as we can  define  higher-level beables which are independent of what is going on at smaller scales, and then   we can  argue that some of the details of physics at smaller scales play a role similar to gauge since no part of physical reality depends on those details.  This would mean that quantum field theory is an `effective field theory' in a very strong sense, because there need not be any well-defined underlying microscopic theory at all.

 Furthermore,  we have noted that unitary quantum mechanics is so successful as a description of the microscopic world that it seems quite unlikely any modification of it at that level could reproduce its success. But on the other hand, there are still plenty of \emph{macroscopic} features of our world which unitary quantum mechanics does not explain very well - even if one puts aside the measurement problem, we have features like the arrow of time, dark matter and dark energy, and indeed even gravity itself, which can be neglected at the scales to which we typically apply quantum mechanics. So if any modification of unitary quantum mechanics is to succeed, there is good reason to think it should  modify quantum mechanics at larger scales, i.e. it should not be microscopically-defined.

An example of such an approach is given by the   consistent histories approach\cite{1996consihit}. In this approach, a `history' is defined as a sequence of orthogonal Hermitian projectors associated with times, each of which can be regarded as representing an `event'\cite{doi:10.1063/1.533050,Isham_1998}. Using a suitable mathematical criterion we can form sets of   `consistent histories' - i.e.  sets of histories such that the sum of probabilities of all members of a set equals one, and all pairs
of histories within the set are orthogonal (so that probabilities add up in the usual way we expect from classical probabilities, without interference effects). Typically the histories appearing in a consistent set are made up of well-decohered quasi-classical variables, which ensures orthogonality.

Now, in some cases the consistent histories approach is just considered as a covariant reformulation of the theory\cite{Anastopoulos_2001}.  However, one can also view the formalism as a putative solution to the measurement problem. Indeed, several different approaches are possible. One could advocate a `many-histories approach' which tells us that we should think of all of the histories as equally real\cite{10.2307/20117986} - in this case the approach could possibly be regarded as unitary-only and indeed looks somewhat akin to the Everett interpretation. One might also consider a version in which exactly one history from every consistent set is real\cite{1996consihit}. But for our present purposes, what matters is the possibility of arriving at a single-world realist consistent histories (SWRCH) approach. To achieve this, we must proceed in two steps: first we must somehow select one consistent set out of all the possible consistent sets, and then we must select a single history from that set in accordance with the associated classical probability distribution, with the chosen history then constituting the actual history of our definite classical world. In this picture the `beables' would be given by the events featuring in the selected history, and all the other parts of the quantum state would be regarded as merely a rule for establishing the correct probabilities for these beables.

Assuming for the moment that we have in hand a satisfactory way of selecting a consistent set, SWRCH  does indeed tell us that our observable reality supervenes on beables which are not microscopically-defined: 
as Sorkin puts it, `\emph{(the consistent histories) picture tends to deny the existence of the microworld. It leads us to identify reality not with an individual history $\gamma$, but with	an element of (the space of macroscopic events), and this would withhold meaning from any statement referring to
	individual atoms or other forms of microscopic matter}'\cite{Sorkin_2007}. Sorkin sees this as a disadvantage, but from our point of view it is a major selling point, because we have already observed that this may be exactly what is needed if we are to have a SWR approach which is capable of accommodating QFT. Indeed, SWRCH provides a helpful case study to show what non-microscopically-defined beables would actually look like. We are so accustomed to thinking of reality as being made up out of microscopic building blocks that the idea of it supervening on beables which are not microscopically-defined may  seem incomprehensible, but SWRCH shows us how such a thing can be done: the history which constitutes our reality includes only events which are sufficiently decohered, so there are  no beables to be found below a certain scale, and thus SWRCH tells us that thus the intuitive picture of spacetime as being filled up with  real pointlike degrees of freedom is  simply  incorrect - the actual physical ontology is defined at a much higher level.

We observe that SWRCH seems to satsify the desiderata set out in section \ref{criteria}: 

\begin{enumerate} 
	
	\item  It preserves the unitary evolution of the quantum state, because the history is defined `on top of' the quantum description and is epiphenomenal - the history selected does not change anything about the unitary quantum description.
	
	\item  It tells us that decoherence plays a crucial role in making histories orthogonal so that they can feature in consistent sets, so decoherence is indeed responsible for the emergence of classical reality 
	
	\item It is  flexible, since it simply instructs us to construct histories out of observables and makes no claim about the ontology underlying those observables, so it does indeed provide a `recipe' which can be applied to any quantum-mechanical system. 
	
	\end{enumerate} 

Since SWRCH avoids some of the major pitfalls which prevent other SWR approaches from successfully accommodating QFT,  it seems very possible that an appropriate formulation of SWRCH could work with QFT. Up until now the consistent histories formalism has mostly been applied within non-relativistic quantum mechanics, but  relativistic generalizations have been developed\cite{Griffiths_2002,hartle2018quantum,osti_466871,BLENCOWE199187,Isham_1998}, and calculational techniques from the consistent histories approach have been applied to  relativistic systems\cite{Anastopoulos_2019}. Indeed,   Sorkin takes the view that the consistent histories approach is actually better off in the QFT context\cite{https://doi.org/10.48550/arxiv.gr-qc/9302018}. Thus for the moment we will optimistically assume that something like SWRCH can indeed reproduce all the empirical predictions of QFT, although more work is needed to establish this formally.

Again, there are some problems to deal with. First, one might worry that we will not be able to select exactly one history from a consistent set in a completely precise way. To do this, we need to be able to determine which events belong to a given history without relying on vague notions like `observer' or `macroscopic,' and it seems natural to do this by requiring that all histories are  maximal - i.e. within a given set, each history contains as many events as possible, conditional on the set remaining consistent. But naturally occurring decoherence is not exact and thus histories made out of natural quasi-classical projection operators will never \emph{exactly} satisfy the consistency conditions used to define consistent sets\cite{DowkerKent}, so it would seem that we have to allow for sets of histories which are merely \emph{approximately} consistent.  And if we allow that, then in many cases there will be no fact of the matter about whether or not a given event belongs to a certain history: although we may not be able to add the event to the history while remaining exactly consistent, we may be able to add it while remaining   approximately consistent, and thus what is included in a given history will end up depending on our arbitrarily chosen cutoff for what counts as `approximately consistent.' This is problematic, as it is  unclear how our observable reality can supervene on a history if there is no non-arbitrary way of saying which events that history contains.

 %  One might hope to make the histories exact by stipulating

That said, ref \cite{1996consihit} presents some examples on finite-dimensional Hilbert spaces and then concludes `\emph{The intuition one can glean from calculations of consistent histories on finite-dimensional Hilbert spaces is admittedly limited, but it does suggest that the solution space of the exact consistency equations is quite large enough for all theoretical purposes so far suggested. In particular, naive but plausible counting arguments suggest that, in the neighbourhood of generic approximately consistent sets of histories, an exactly consistent set can be found.}' Roughly speaking the argument is that in a large enough Hilbert space, for any approximately consistent set there will be an exactly consistent set in its very near neighbourhood, so for any set of histories made out of quasiclassical projection operators we should be able to find another very close  set which is exactly consistent. If this is true, we can in fact get precisely defined histories out of the decoherence process after all, since we are free to choose the preferred set in such a way that it is exactly consistent. We reinforce that this example covers only the finite-dimensional case for non-relativistic quantum mechanics, and the considerations adduced in ref \cite{pittphilsci20537} stand as a warning against assuming that approaches which work for  non-relativistic quantum mechanics will transfer over to the case of quantum field theory, so  although these results offer grounds for cautious optimism, no firm conclusions can be drawn until the results are extended to full quantum field theory.  However, it is nonetheless interesting to see that even though decoherence is approximate and inexact,  something precisely defined can apparently emerge from it. In addition to the importance of this fact for the interpretation of quantum mechanics, it may have interesting consequences for the philosophy of emergence. 

That said, even if we can get precisely defined histories from this formalism we still have problems - in particular, difficulties associated with the stability of the histories  may undermine the claim of SWRCH to be a SWR approach in the first place. For even if a history belonging to a consistent set is quasi-classical up until the present day this does not mean it will be quasi-classical into the future, so we are left with the somewhat unsettling prediction that classical reality could simply break down at any moment\cite{1996consihit}. Even more problematically, records in histories belonging to consistent sets are not   stable -  if a history belonging to a consistent set contains records at some time $t$, there is no   reason to expect that those records reflect events that actually occurred earlier in the history\cite{1996consihit}. Thus some of the epistemic problems discussed in section \ref{unitary} may creep back in: observers in a world as defined by SWRCH wouldn't be able to take for granted that  records and memories are informing them accurately about the measurement outcomes that occurred in the past, so the process of empirical confirmation would be undermined. 

Of course, in principle we can solve these problems by simply stipulating that the `preferred set' is one in which the histories are quasi-classical and in addition it is a  `strongly decoherent' set, meaning that in each of the histories generalized records accurately reflecting the past are preserved at every step\cite{https://doi.org/10.48550/arxiv.gr-qc/9509054}. But this seems  somewhat ad hoc - we have no well-motivated model grounding the stability of the histories, we are simply stipulating that the set must be strongly decoherent because  we want to be able to say that our memories of the past can be relied upon. Of course we should not expect a scientific theory to \emph{prove} that our memories of the past are accurate, but ideally we would like the fact that memories and records are mostly accurate to emerge in a reasonably natural way from an interpretation or modificatory strategy for quantum mechanics  -   if it simply has to be stipulated as an add-on to the interpretation, one may worry that this is not enough to justify our usual practices of empirical confirmation.  After all, the Everettians  might quite reasonably object that the ad hoc assumption that the history of the actual world is selected from a strongly decoherent consistent set is no better than  (or indeed - \emph{worse} than) their own  assumption that we are inside a branch of the wavefunction which is associated with a high mod-squared amplitude.

Various attempts have been made to offer a less ad hoc way of selecting an appropriate preferred set, but it's not clear that any of them is fully satisfactory if we are aiming for a SWR approach. For example,    Gellman and Hartle\cite{GellmanHartle} have suggested an approach based on the observation that the kinds of beings that we are (IGUSes - information gathering and utilising systems) can only exist in sufficiently quasi-classical histories, so our consciousness will necessarily divulge to us an appropriate quasi-classical history from a set containing such histories. However, it's implausible to think that this approach will yield \emph{exactly one} set of consistent histories, so this would not qualify as a SWR approach and therefore it would presumably suffer from exactly the problems we have already mentioned in connection with non-SWR approaches: for example, Dowker and Kent point out that `\emph{there is, in this interpretation, no correlation between the experiences of these splendidly isolated IGUSes; each may well believe itself in communication with others, but the others may be experiencing a quite different history, or nothing at all.}'\cite{1996consihit}   Gell-Man and Hartle\cite{GellmanHartle} also separately suggest that there might turn out to be one set which is much more `quasiclassical' then the other sets, and this one is by definition the one representing our actual quasiclassical domain;  similarly Anastopoulos\cite{Anastopoulos_2019} argues that we should choose a preferred set such that `\emph{each consistent subset of it can be embedded into the lattice of propositions of a classical history theory.}' These ideas seem quite reasonable within a many-histories approach where all the sets are physically real, because then we can plausibly regard the choice of a consistent set as essentially just a representative choice - they all correspond to the same physical reality, and the preferred set is preferred simply insofar as it is more perspicacious for understanding the relation between  physical reality and our quasiclassical experiences.  But if the intention is to probabilistically select a single history from the preferred set this won't work:  the single history selected from the preferred set will not map to a unique history in all of the other sets, and thus the choice of preferred set is not just a representational choice but a physically substantive stipulation which plays a major part in determining the contents of reality. Surely this stipulation must be made on some physical grounds, rather than merely by consideration of which description is more convenient for us.

Nonetheless, although SWRCH doesn't currently offer a complete solution to the measurement problem, it is important as a template - for in  light of  Wallace's argument, we think it is likely that any precisely defined SWR which can reproduce QFT must look something like SWRCH, albeit supplemented in certain ways.   Thus this discussion helps  focus attention on what remains to be done: we  need some principled way of choosing a preferred set of histories which ensures it is very likely to be a quasi-classical, strongly-decoherent set.

\section{The Non-Dynamical Approach} 

The final route we will explore involves allowing that the unique real world associated with a SWR approach could supervene on beables which are not dynamical. The idea of reality supervening on beables which are not dynamical may seem counterintuitive, but in fact there are a number of indications that we should be taking non-dynamical approaches to physics seriously. The standard `dynamical' picture, where the universe   can be regarded as something like a computer which takes in an initial state and evolves it forward in time\cite{Wharton}, was bequeathed to us by Newton; but physics has changed a great deal since Newton's time, and many of our modern theories don't fit naturally into this dynamical picture. For example,  the solution to the Einstein equations of General Relativity is not a state at a time but an entire history of a universe\cite{wald2010general}, so it doesn't seem very compatible with the dynamical time evolution picture\footnote{A time-evolution formulation of the Einstein equations does exist\cite{ringstrom2009cauchy, foures1952theoreme}, but the original global formulation remains central to research in the field and there seems no obvious reason to think that the time-evolution formulation must be more fundamental.};  many more examples are presented in refs  \cite{adlam2021laws, chen2021governing,Wharton,Deutsch_2015}. Thus, since it seems quite possible that we will have to move away from the dynamical picture in any case, it makes sense to consider whether non-dynamical approaches could help solve the measurement problem.

Roughly speaking,  a non-dynamical solution to the measurement problem would involve extracting from the unitary evolution of the quantum state a set of classical histories together with a probability distribution over those histories which at least approximately reproduces the probabilities assigned by unitary quantum mechanics. Then we need only postulate something like a single, atemporal collapse of the wavefunction which selects a unique classical history according to the probability distribution prescribed by the entire dynamical history of the quantum state. In principle this approach could be thought of as simply the Everett interpretation plus an additional step in which   exactly one of the Everettian histories is probabilistically selected and actualised, and thus in principle it should have all the  advantages of the Everett approach whilst also  avoiding the epistemic problems associated with non-SWR models - in particular, it would be  empirically equivalent to the Everett interpretation, or rather to a version of the Everett interpretation in which the mod-squared amplitudes can be given a probabilistic interpretation, and thus it would be able to make all the same empirical predictions as the Everett interpretation, including reproducing all of QFT. But it is not trivial to specify this approach in a precise way, for the     Everettian histories become distinct via the mechanism of decoherence, so they are only approximate and emergent, and therefore we need to add something further to define the set of histories from which we will be selecting exactly one. 

Obviously, one such approach as SWRCH - it is in fact both non-microscopically-defined and also non-dynamical, since to determine whether a set of histories is consistent we must apply the appropriate mathematical criterion to the entire histories, so we can't think of the histories as being generated in a step-by-step dynamical process. For example, a pair of histories which decohere up to some point in  $t$ can have their decoherence destroyed by later events which cause the original alternatives to interfere, so the set of available histories at a time $t$ does not depend only on events up until time $t$\cite{Sorkin_1997_b}. However we have seen that SWRCH faces some significant difficulties, so let us see if there is an alternative non-dynamical approach which can overcome some of these issues. 

Consider    Kent's `Lorentzian solution to the quantum reality problem'\cite{Kent,2015KentL,2017Kent}.  Kent suggests a model in which the wavefunction undergoes its usual unitary evolution until the end of time, and   a measurement is performed on the final state, and the actual course of history is determined by that measurement.  For example ref \cite{2017Kent} suggests a final measurement of the distribution of photons across spacetime, and then stipulates that the actual contents of reality (i.e. beables) are given by a mass-energy distribution over spacetime such that the value at a given point x is equal to the expectation value of the stress-energy tensor at x, conditional on the outcome of the final measurement outside the future lightcone of x\footnote{Kent also makes allowance for the possibility that there is no end of time - in this case we simply take a limit as $t \rightarrow \infty$, making some assumptions which ensure that the limit is well-defined.}. In Kent's approach it is indeed the case that our observable reality supervenes on beables which are non-dynamical, because his beables  don't evolve, but rather they are determined   all at once from the result of the final measurement. Indeed, Kent's beables are purely epiphenomenal, since the evolution of the quantum state is completed before the beables are chosen, and  the beables are determined only by the result of the measurement on the final quantum state, so the beables at one spacetime point have no direct dependence on the beables at other spacetime points.

 Kent's approach offers solutions to some of the problems we raised previously for SWRCH. It provides a natural way to select one particular set of consistent histories, i.e. those which are recorded in the final state in the position basis. Moreover, these histories are indeed precisely defined: each history corresponds to one possible result of the final measurement and includes exactly the set of all events which are recorded in the result of the measurement. And finally, Kent's mechanism singles out a set of histories which are guaranteed to contain records accurately reflecting the past,   because in this picture an event can occur only if it is recorded in the final state, so  by definition the history defined from the result of the final measurement must have the property that accurate records of every event which occurs persist until the final measurement. Thus the stability of the macroscopic world and the reliability of records is built into Kent's ontology in a non ad-hoc way. 
 
In addition, Kent's approach seems to satisfy the desiderata of section \ref{criteria}: 

\begin{enumerate} 
	
	\item It allows the unitary quantum evolution to proceed unimpeded up until the end of time, because the history selected in the final measurement is selected after the evolution is finished and hence is epiphenomenal. 
	
	\item It allows us to attribute the emergence of classicality to decoherence - indeed, Butterfield characterises Kent's approach as a decoherence-based approach to the measurement problem which gets the role of decoherence right, since it does not `\emph{make the error of thinking that an improper mixture is ignorance-interpretable}'\cite{Butterfield_2019}.  The function of decoherence in  Kent's approach is to produce  a set of quasi-classical histories, including all the macroscopic events we would normally expect to feature in a description of reality; the final measurement simply defines a preferred basis and then selects one of the classical histories produced by decoherence in that basis.  
	
	\item It is  quite flexible - for any given quantum description, we can ask whether or not some event occurring in that description will be recorded in the state of the world at the end of time, and thus in principle the prescription can be applied to any scenario which can be described by quantum mechanics.
	
	\end{enumerate} 

As with the consistent histories approach, Kent's approach avoids the obvious pitfalls when it comes to accommodating QFT and thus it seems possible that this approach could be made to work in the QFT domain. Indeed, it was explicitly formulated to be Lorentz-covariant, so it is appropriate for relativistic quantum mechanics. Although it does not yet have a fully formulated QFT version, we can say roughly how such an approach would work: whenever a quantum field theory experiment is performed there is inevitably  a macroscopic record of the result (even if that record is just the state of the experimenter's brain), and that record is left in a quantum  state $\rho$, which is calculated using the standard methods of QFT. Since the record is macroscopic, we can assume that $\rho$ takes the form of an improper mixture: $
\rho  = \sum_i c_i |  i \rangle \langle i |$ for some set of orthogonal states $ |  i \rangle$ representing the possible measurement results and some nonnegative coefficients $c_i$ which sum to one. Macroscopic records will reliably leave traces in the final state, so we can assume that the final state takes a corresponding form $
\rho' = \sum_i c_i |  i \rangle_T \langle i |_T$, where the set of orthogonal states $ |  i \rangle_T$ are the appropriate time-evolved analogues of the states $ |  i \rangle$, the coefficients $c_i$ remain the same, and we trace over parts of the final state irrelevant to this measurement. Then Kent's approach predicts that in this QFT experiment the probability for obtaining the outcome associated with state  $ |  i \rangle$ is equal to the probability of obtaining the outcome associated with state $  |  i \rangle_T$ in the final measurement, which is given by $c_i$. But by construction $c_i$ is also the probability  predicted for this outcome by the usual  methods of quantum field theory, so Kent's approach should in principle successfully reproduce the  empirical predictions of quantum field theory. However, we  note that these predictions must be understood as applying directly to the macroscopic records of experimental outcomes - they cannot be understood as pertaining directly to the behaviour of fundamental particles or fields, since particles and fields are not the beables of Kent's approach. This may seem surprising but it should be understood in the context of Wallace's criticisms, which emphasize that attempting to postulate some specific underlying ontology for quantum field theory doesn't work very well - so it is perhaps not so surprising that an ontological approach which can reproduce the predictions of QFT makes predictions for  macroscopic events directly while refraining from associating any physical ontology with most of the underlying structure of QFT.

However, it should be noted that Kent's explicit models for this approach characterise the measurement at the end of time as a measurement on the positions of a set of particles   - ref  \cite{Kent}.  uses a collection of bosons and fermions, and refs \cite{2015KentL,2017Kent} use a collection of photons. Thus since particles are not fundamental in QFT, one might worry that Kent's approach faces similar problems to the other particle-based approaches we have examined. However there is a crucial difference between Kent's approach and something like de Broglie-Bohm, because in Kent's approach our observable reality supervenes not on the particle positions  but on the result of a measurement performed on the particle positions, and therefore arguably Kent doesn't need the particles to be fundamental or precisely-defined, since we can get a precise  result when we measure the position of a particle even if the quantum state of that particle is  smeared across space. Moreover, the measurement occurs only at the end of time, so the particles only need to exist asymptotically - they need not even be approximately defined at other times. And since QFT does indeed allow us to make asymptotic position measurements on particles produced during experiments,  it seems reasonable to think that some such procedure can be specified in a well-defined way.

Before moving on, let us briefly pause to consider an objection to the consistent histories framework due to Okon and Sudarsky\cite{https://doi.org/10.48550/arxiv.1309.0792,https://doi.org/10.48550/arxiv.1504.03231}, who observe that in this framework, `\emph{in order to successfully apply the formalism to a
	concrete measurement situation, one needs to know in advance  ... what it is that the apparatus one is using actually measures.}' Thus they argue that the consistent histories approach does not really solve the measurement problem, since in their view a solution to the measurement problem must allow us to give a self-contained description of an individual experimental setup without any external input. Now, this objection is partly pointing to the fact  that the consistent histories formalism does not tell us how to select one particular consistent set, since a choice of consistent set is required to determined the basis of measurement for the apparatus. However, this issue is resolved within Kent's approach, and yet it seems that the same objection still applies there:  within Kent's framework we can't  give a self-contained description of an individual experimental setup, since we have to take into account the entire future history\footnote{Although for all practical purposes we can assume that macroscopic measurement outcomes will be recorded in the final state and thus the dependence on the future doesn't get in the way of applying the framework to real experiments.}. In our view, this shows that there is a problem with the criterion employed by Okon and Sudarsky - insisting on self-contained descriptions of individual experiments takes for granted that the solution to the measurement problem must be a dynamical one in which   reality can be decomposed into autonomous steps of evolution, but in models which are not dynamical, like SWRCH and Kent's approach, it will not always be the case that the theory can describe individual measurement scenarios in isolation, because what happens in one measurement scenario may depend in some way on the rest of the history. Thus since we have argued that there are good reasons to think that the correct solution to the measurement problem will be non-dynamical,  we must reject  Okon and Sudarsky's formulation of the measurement problem. 
	
	\subsection{Superfluous Structure?} 

Suppose for the moment that Kent's approach or some similar non-dynamical approach can succeed in producing a suitable set of precisely defined histories from which we can subsequently select exactly one. We envision an Everettian response which goes something like this: `\emph{in order to arrive at your set of histories you first have to postulate an entire Everettian multiverse from which you subsequently extract the histories. So you are actually committed to the entire Everettian multiverse in any case, and thus your approach is essentially just Everett in denial.'}' A similar objection was made against the de Broglie-Bohm approach by Brown and Wallace\cite{R_Brown_2005}. 

However, this response mischaracterizes the role of the `Everettian multiverse' in this kind of approach. While the process by which the set of histories is constructed would most likely invoke a piece of mathematical structure which looks like an Everettian multiverse, that piece of mathematics is not postulated as part of physical reality; indeed it is not regarded as the kind of thing which \emph{could} constitute physical reality. It is simply a rule for assigning probabilities to histories, and thus assuming that all the branches are physically real is the same kind of mistake as assuming that in classical physics all the dynamically possible histories are physically real.

It may be useful to compare to the simpler case of an indeterministic classical theory. A non-dynamical formulation of such a theory would require two steps. We would start from a  set of `dynamically possible branching histories,' each of which corresponds to a different initial condition and records all the ways the course of history could go after that initial condition. The first step would be  to choose one of these branching histories, which is equivalent to selecting an initial condition; the second step would be to select exactly one of the non-branching histories encoded in this branching history according to the probability distribution prescribed by the theory. Put this way, it is evident that a SWR approach of the kind we have described here is precisely analogous: first we make an arbitrary selection of a possible Everettian multiverse (equivalent to selecting an initial quantum state), and then we  apply some rule, such as Kent's final measurement algorithm, to probabilistically select exactly one of the classical histories encoded in this multiverse\footnote{The need for two selection steps is admittedly somewhat clunky. It would be preferable to combine these steps in some way, or perhaps we could eliminate the first step altogether by invoking something like Chen's quantum Wentaculus\cite{https://doi.org/10.48550/arxiv.2203.02886}. This approach prescribes a unique initial state, so the laws of nature single out a unique multiverse. Then just as in the Newtonian case we would have a single selection step, except that this step would be probabilistic rather than arbitrary.}. So the modal status of   `the other Everettian histories' in this putative SWR approach is the same as     `the other histories in the same branching history' in the classical indeterministic case, and nobody seems inclined to argue that the  other histories   must be physically real in the classical indeterministic case.

In response to this,  the Everettian might argue  that unlike in the classical deterministic case, the branches of the wavefunction interact with each other, so they must be taken as physically real. However, it is wrong to say that the branches are interacting in a non-dynamical SWR approach. In this picture the wavefunction simply provides a rule for assigning probabilities to histories (conditional on a particular choice of initial state): mathematically, this rule has some wavelike features, including sums over complex numbers which look superficially like interference of branches if one thinks of the wavefunction as dynamically unfolding, but  since this picture is explicitly non-dynamical there is not really any dynamical unfolding.  We simply have a rule which ultimately yields some probability distribution over entire courses of history at once, and there is no need for the other branches to be physically real  to justify this mathematical rule.

\section{The Space of Possibilities \label{both}}

  RQM with cross-perspective links,  SWRCH, and Kent's approach provide helpful examples of solutions to the measurement problem which might have a better chance of working for QFT, but they don't necessarily exhaust the options - there have been certainly other proposed solutions to the measurement problem which are not precisely defined, not microscopically-defined, or non-dynamical. Since the field is too large for us to survey comprehensively, we will instead   try to map out   the space of possibilities   by considering the relationship between non-precisely-defined, non-dynamical and non-reductionist approaches.

First, can there be approaches which are non-microscopically-defined but still dynamical? We don't know of any examples, and indeed this seems like quite a difficult thing to achieve in a relativistic context. For if your beables are not microscopically defined, then they presumably have finite spatial extent, and therefore if you want them to undergo temporal evolution in the usual sense you will   need to select a preferred reference frame on which the evolution takes place, since all the parts of your spatially extended beables will have to undergo each step of evolution at the same time - you can't simply break your beables up into smaller pieces and evolve the parts separately, because that would amount to returning to microscopically-defined beables. And although it's possible for a theory with a preferred reference frame to be empirically compatible with relativity a if  the preferred frame can't be inferred from the empirical data, nonetheless a preferred reference frame is an uncomfortable fit with relativity's denial of absolute simultaneity, and therefore most physicists would prefer to avoid such a thing.   

In addition, it seems quite difficult to formulate a non-microscopically-defined  dynamical  approach which guarantees stability of the quasi-classical regime into the future. This can be seen from the fact that consistent sets  generically have the feature that histories which are quasi-classical up to some point in time can cease to be quasi-classical thereafter\cite{1996consihit}. Sorkin\cite{Sorkinpath} notes that `\emph{To avoid this one would have to consider sets of histories with temporal support reaching arbitrarily far into the future. This would entail a highly uncausal prediction algorithm,}' and concludes that solutions to this problem `\emph{ are mainly of a “teleological” type.}' From this point of view, we can see that it is not a coincidence that Kent's model puts the selection of the course of history into the distant future:   it is precisely this which allows the approach to assure us in a non-ad-hoc way that all the resulting histories will be strongly decoherent, so that memories and records are generally accurate reflections of this past.  Thus we consider it likely that a viable solution to the measurement problem which is non-microscopically defined will also be at least to some degree non-dynamical. 

On the other hand, it does seem possible to have approaches which are non-dynamical but still  microscopically-defined. Indeed, Kent's approach is of this kind - we have seen that Kent's beables are non-dynamical, but they are still microscopically-defined since Kent's framework defines beables at individual spacetime points. Thus  Kent's approach does in fact provide us with a `primitive ontology' which lives in 3D spacetime. That said, one might argue that Kent's beables are not really playing the role envisioned by the proponents of primitive ontology, because they are non-dynamical and epiphenomenal and thus are not really `building blocks' in the usual sense of that term - although our observable reality supervenes on the beables, the beables at a given time depend on facts about the future evolution and thus this approach doesn't  allow us to break phenomena down into  the autonomous behaviour of beables at individual spacetime points. Indeed,  Kent's beables are  `microscopically-defined' in quite an idiosyncratic way, for many of the events that we usually expect to be part of the microscopic regime, such as a particle passing through one slit or another in a double-slit experiment, will not leave traces in the final state and thus will not be represented in the beables defined by the final measurement; yet macroscopic events, such as a measuring device registering one result or another, are virtually guaranteed to leave traces in the final state, and thus they will reliably be represented in the beables defined by the final measurement. So although the beables produced by Kent's prescription are defined at individual spacetime points,   they are nonetheless more likely to be associated with what we usually regard as macroscopic variables   than the kinds of microscopic variables appearing in standard quantum mechanics and QFT. Indeed, this is precisely how Kent's approach avoids the problem posed by Wallace: we are not required to find a quasi-classical variable inside the highly non-classical regime of QFT, because that regime need not be represented directly in the beables at all - it influences the beables only indirectly via its role in assigning probabilities to quasi-classical histories. 

This suggests to us that the  `microscopically-defined' character of Kent's model may not persist in subsequent formulations. For the crucial feature of Kent's approach is that `the course of history' is determined by the measurement on the final state - this leaves room for different possible methods of mapping this measurement result to an actual history, and therefore the actual history does not necessarily have to be understood in terms of an ontology of properties defined at individual spacetime points. Indeed the level of resolution used by the current models seems superfluous:  most of the events recorded in these beables will be macroscopic rather than microscopic, so the beables could represent exactly the same macroscopic events, and hence produce exactly the same conscious experiences, were they defined in a more coarse-grained way. So although Kent's approach shows that it is possible to have a model which is non-dynamical but microscopic, we think it is likely that other possible solutions in this domain will be both non-dynamical and non-microscopically-defined. 

  Finally, we note that   SWRCH and Kent's approach demonstrate that it's possible to have models which are non-reductionist and/or non-dynamical but still precisely defined. On the other hand, one can certainly imagine models which are  non-reductionist and/or non-dynamical and \emph{also} not precisely defined, and this might be an interesting possibility to explore further.

 \section{Conclusion}
 
Although there is work to be done on all of the approaches we have described here, we think   the outlook is positive:   once quantum field theory and epistemic issues (and perhaps also quantum gravity - see ref \cite{https://doi.org/10.48550/arxiv.2204.08064}) are taken into account, the prospects for settling on a definitive solution to the measurement problem seem quite good. In particular, we think it is likely that a successful interpretation  or modificatory strategy will have the following four features: 
 
 \begin{itemize}

 	\item It makes no substantial changes to the formalism of unitary quantum mechanics (at least at the microscopic level)
 	 	
 	\item Decoherence plays a significant role in the emergence of classical reality
 	
 	\item Observers (approximately) see a unique outcome to each measurement and are able to (approximately) establish a shared observable reality
 	
 	\item This shared observable reality supervenes on beables which are approximate and emergent, and/or non-dynamical, and/or non-microscopically-defined

 	\end{itemize}

 As we have seen, the possibilities for approaches which combine all of these criteria are quite limited, which    means there is some hope of actually reaching a consensus on this issue. Of course, all of the approaches we have singled out as promising remain quite under-developed; Wallace very reasonably argues that we should judge on the basis of the theories we actually have, not the theories we hope to have one day\cite{pittphilsci20537}, and it's true that  RQM with cross-perspective links, SWRCH and Kent's Lorentzian classical reality approach are in some ways more promissory notes than complete theories. But if it is accepted that the unitary-only approaches are untenable, then it would seem  these promissory notes may be the only remaining possibility. We don't necessarily think that any one of the approaches we have described here represents the final solution to the measurement problem, but they exemplify features which seem likely to appear in the final solution and thus they provide a useful starting point for further investigation.

 There is also a higher-level conclusion to be drawn here about strategies for solving the measurement problem.  Many  solutions to the measurement problem  proceed by trying to make quantum mechanics look more classical and hence more palatable to our classical intuitions - so for example the `primitive ontology' approaches aim to reduce everything down to fundamental building blocks living in spacetime, just like the physical objects that we are familiar with.  And indeed, making quantum mechanics look more classical is  a good strategy if the primary aim is for us to feel that we understand quantum mechanics. But the measurement problem is not just about our desire to have a subjective feeling of understanding: it is a question about the nature of our reality which presumably has an objectively correct answer, and making quantum mechanics look more classical is not necessarily a good strategy if the aim is to arrive at that objectively correct answer. For after all, surely at least part of our difficulty in making sense of quantum mechanics is due to the fact that we are, inevitably, trying to understand it through the lens of our classical intuitions, and thus we are  inadvertently imposing on it classical assumptions which may not really be a good fit for the physics. If that is so, the right approach is actually  to try to make quantum mechanics \emph{less} classical - that is, we should figure out which classical assumptions we have  imposed on our thinking about quantum mechanics, and decide which of those assumptions should be taken away. 

That is exactly what we have seen in this article. For we have argued that there are good reasons to think that in order to solve the measurement problem, at least one of the following classical intuitions must be given up: a) physical reality is fundamentally dynamical, b) our observable reality supervenes on the behaviour of something microscopically-defined, c)  our observable reality supervenes on something precisely defined. So  in our view it is likely that the measurement problem will ultimately be solved not by cleaving to our classical intuitions but by venturing further away from the classical picture of the world.

 \section{Acknowledgements} 
 
 Thanks to Carlo Rovelli and Adrian Kent for some interesting discussions on topics covered in this paper.  This publication was made possible through the support of the ID 61466 grant from the John Templeton Foundation, as part of the “The Quantum Information Structure of Spacetime (QISS)” Project (qiss.fr). The opinions expressed in this publication are those of the author  and do not necessarily reflect the views of the John Templeton Foundation.

   \bibliographystyle{unsrt}
 \bibliography{newlibrary12}{}

 \end{document}